\documentclass[12pt]{article}

\ifx\pdfoutput\undefined
\usepackage[dvips,bookmarks]{hyperref}
\else
\usepackage{hyperref}
\fi
\hypersetup{colorlinks=false,bookmarksopen,bookmarksnumbered,citecolor=blue,
   pdfstartview=FitH}

\usepackage[top=3cm, bottom=3.25cm, left=2.75cm, right=2.75cm]{geometry}
\usepackage{graphicx}
\usepackage{amsfonts}
\usepackage{amsthm}
\usepackage{amsmath}
\usepackage{amssymb}
\usepackage{epstopdf}
\usepackage{tikz} 
\newcommand{\beq}{\begin{equation}}
\newcommand{\eeq}[1]{\label{#1}\end{equation}}
\newcommand{\ber}{\begin{eqnarray}}
\newcommand{\eer}[1]{\label{#1}\end{eqnarray}}
\newcommand{\non}{\nonumber}
\newcommand{\ba}{\begin{array}}
\newcommand{\ea}{\end{array}}
\newcommand{\re}[1]{(\ref{#1})}

\numberwithin{equation}{section}

\newcommand {\cK}{{\cal K}}
\newcommand {\cL}{{\cal L}}

\newcommand {\cS}{{\cal S}}

\newcommand{\pa}{\partial}                           

\newcommand{\vf}{\varphi}

\def\a{\alpha}
\def\b{\beta}

\def\d{\delta}
\def\e{\epsilon}
\def\f{\phi}
\def\g{\gamma}

\def\m{\mu}

\def\s{\sigma}

\def\L{\Lambda}

\begin{document}
\begin{titlepage}
\vspace{5mm}

\begin{center}
{\Large \bf Higher Spin Versus Renormalization Group Equations
}
\\ 
\end{center}
\vspace{5mm}
\begin{center}
{\bf Ivo Sachs} \\
\vspace{5mm}

\footnotesize{
{\it Arnold Sommerfeld Center for Theoretical Physics\\ Ludwig-Maximilians University\\
Theresienstr. 37, D-80333 M\"unchen\\
Germany}\\ivo.sachs(at)lmu.de}

\end{center}
%
%
%
%


\vspace{10mm}
\abstract{We present a variation of  earlier attempts to relate renormalization group equations to higher spin equations. We work with a scalar field theory in 3 dimensions. In this case we show that the classical renormalization group equation is a variant of the Vasiliev higher spin  equations with Kleinians on AdS$_4$ for a certain subset of couplings. In the large N limit this equivalence extends to the quantum theory away from the conformal fixed points. }

\end{titlepage}

\section{Introduction}  
In view of possible generalizations of the AdS/CFT correspondence away from  conformal fixed points and, perhaps more importantly, deriving it from field theory, an interesting idea is to think of it as a geometric realization of the renormalization group (RG) flow (see eg. \cite{Akhmedov:1998vf,Alvarez:1998wr,de Boer:1999xf,Akhmedov:2010sw,Heemskerk:2010hk,Litim:2011qf} for a various attempts in this direction). One strategy, \cite{Das:2003vw,Douglas:2010rc} (and more recently, \cite{Zayas:2013qda}) is to relate higher spin (HS) equations to the RG-equation for a non-local mass term in a free field theory. 

In this note we consider a simpler, more restricted model, consisting of a scalar field theory in 2+1 dimensions. After a short review of Polchinski's form \cite{Polchinski:1983gv} of the Wilsonian RG-equation we then consider the RG-flow for local, quadratic, higher derivative couplings. We first map the linearized RG-flow of these couplings, as functions of the cut-off scale to the  {\it non-propagating, auxiliary or topological} sector of the free, non-minimal Vasiliev-type theory, with outer Kleinians  \cite{Vasiliev:1992av} (and  \cite{Vasiliev:1999ba} for a review on HS theories) on a four-dimensional AdS-background. This mapping turns out to be a straight forward consequence of the simple relation between the 3- and 4-dimensional HS-algebras in the twistor representation \cite{Vasiliev:2012vf}. 

Next we consider the representation of  HS-gauge transformations on the RG-equations. While we are not able give an interpretation of generic  HS-gauge transformations we will analyze some simple examples in detail. In particular, we will see that while four dimensional HS-gauge transformation do not leave the field theory action invariant in general, the RG-equation transforms covariantly with respect to such transformations. 

Finally, we describe how non-linear terms in the RG-flow affect the HS-equation of motion. At the classical level these interactions produce an inhomogeneity in the HS-equation which nevertheless preserves HS-gauge invariance. The reason for this to work is that while  the non-linear flow induces a plethora of other couplings in addition to the higher spin couplings described so far these extra couplings do not effect the RG-flow of the latter. At quantum level this is no longer the case. Nevertheless, we show that in the usual large N limit of an N-component scalar field theory the classical HS system is merely affected by anomalous dimensions which enter as a HS-singlet in the RG-equation. 

Finally we comment on possible applications of our findings to the HS-AdS/CFT duality  at large N \cite{Klebanov:2002ja,Giombi:2009wh}.

\section{Polchinski Equation}
We start with a single scalar field in $d=2+1$ dimensions. Following the conventions of  \cite{Osborn:2011kw} the Polchinski equation, expressed in terms of dimensionless coordinates, momenta and fields takes the form
\ber
\left(\frac{\partial}{\partial t}+ D\varphi\cdot\frac{\d}{\d\vf}\right)\cS_t[\vf]&=&\frac{1}{2}\frac{\d}{\d\vf}
\cS_t[\vf]\cdot G\cdot \frac{\d}{\d\vf} \cS_t[\vf]-\frac{1}{2}\frac{\d}{\d\vf} G\cdot \frac{\d}{\d\vf} \cS_t[\vf]\,.
\eer{p1}
Here, $\cS_t[\vf]$  is basically  the Wilsonian effective action, minus the kinetic term, at cut-off scale $t=-\ln(\L/\L_0)$, expressed in terms of the couplings and dimensionless fields $\L^{\d}\vf(x\L):=\f(x)$.  $\cS_t[\vf]$ is essentially local, meaning that the non-locality has an all-orders Taylor expansion for small $p^2$ (see e.g. \cite{Rosten:2010vm}). The UV-regularized propagator, $\cK(p)/p^2$,  enters through $G(p^2)=2\cK'(p^2)$ with $\cK(0)=1$, $\lim_{p^2\to\infty} \cK(p^2)=0$. For simplicity we will also assume that $\cK(p^2) $ is analytic in the neighborhood of the real positive semi-axis with essentially exponential fall-off at infinity. Finally, $D$ is the dilatation operator which acts on $\varphi$ as
\beq
D\vf(p)=-(p\partial_p+d-\d)\vf(p)\,,
\eeq{dildef}
where $\d$ is the canonical dimension of $\vf$. Expanding $\cS_t$ in the fields and couplings we can, in principle, extract the renormalization group equations. To illustrate this we set the right hand side of \re{p1} to zero which is justified in the linearized approximation for a quadratic action and neglecting the cosmological constant. For instance, for $\cS_t=m^2\int \vf(p)\vf(-p)$ eqn \re{p1} then gives
\beq
\left(\partial_t m^2-2m^2(d-\d)\right) \int \vf(p)\vf(-p) -m^2\int \left(p\partial_p\vf(p)\vf(-p)+\vf(p)p\partial_p\vf(-p)\right)=0\,.
\eeq{}
Using $\d=\frac{d-2}{2}$ for a scalar with canonical kinetic term and performing a partial integration on the last term we recover
$\partial_t m^2-2m^2=0$. The marginal deformation in $2+1$ dimensions is $\cS_t=\lambda \int \vf(p_1)\cdots\vf(p_6)\delta^3(p_1+\cdots+p_6)$ with $D\varphi\cdot\frac{\d}{\d\vf}\cS_t=2(d-3)\cS_t$. We should note that in this case the second (quantum) term on the r.h.s. of  \re{p1} acts as a source term for lower order couplings. Similarly, for $\cS_t=g^{ab}\int p_a\vf(p)p_b\vf(-p)$ we get 
\ber
&&\left(\partial_tg^{ab} -2g^{ab}(d-\d)\right)\int p_a\vf(p)p_b\vf(-p) \\&&\qquad\qquad-g^{ab}\int \left(p_ap\partial_p\vf(p)p_b\vf(-p)+p_a\vf(p)p_bp\partial_p\vf(-p)\right)=0\non\,,
\eer{}
which, upon partial integration amounts to $\partial_tg^{ab}=0$. More generally, for interactions of the form 
\beq
\cS_t=g^{a_1\cdots a_n}\int p_{a_1}\cdots p_{a_n}\vf(p)\vf(-p)\;,\qquad  g^{a_1\cdots a_n}=g^{(a_1\cdots a_n)} 
\eeq{sc1}
This is a special case of the more general HS coupling analyzed in \cite{Bekaert:2010ky}. Unlike there, we assume the couplings to be independent of the coordinates of the coordinates of the 2+1 dimensional field theory because we want to apply standard Wilsonian methods for the RG-flow. Eqn \re{p1} then yields at linear order
\beq
\left(\partial_t   g^{a_1\cdots a_n}+(n-2)g^{a_1\cdots a_n}\right) \frac{\partial}{\partial g^{a_1\cdots a_n}} \cS_t[g_t,\vf]=0\;.
\eeq{grd}
As long as $\cS_t$ is quadratic in $\vf$ this is exact in the linearized approximation up to the cosmological constant $\L$, due to the second term on the r.h.s. of \re{p1}. $\L$ can receive a contribution proportional to $g^{a_1\cdots a_n}$ but obviously does not affect the running of  $g^{a_1\cdots a_n}$. 

In what follows we shall concentrate on the RG-flow of the {\it traceless} subset of couplings of the form \re{sc1}, first in linearized and classical approximation. In this case we will show that the RG-equation \re{p1} is actually a free HS-equation on AdS${}_4$. 
Following \cite{Vasiliev:2012vf} we note that the dilatation operator $D$ which is part of the   conformal algebra in $2+1$ dimensions has a Weyl star-product realization in terms of quadratic products of  $(y^-_\a,y^{+\a})$ satisfying $[y^-_\a,y^{+\b}]_*=\d_\a^\b$. In particular, 
\beq
 D=\frac{1}{2}y^{+\a}y^-_\a \qquad\hbox{and }\qquad P_{\a\b}=iy^-_\a y^-_\b\,, 
\eeq{pd}
where $P_{\a\b}$ is the translation generator, whereas 
\beq
L^\a_{\;\;\b}=y^{+\a} y^-_\b
-\frac{1}{2}\d^\a_\b y^{+\g} y^-_\g\,,\qquad\hbox{and}\qquad  K_{\a\b}=-iy^{+}_\a y^+_\b
\eeq{ls}
represent Lorentz transformations and special conformal transformations respectively. We then replace $g^{a_1\cdots a_n}(t)p_{a_1}\cdots p_{a_n}$ by 
\beq
g^{(n)}(t)\equiv e^{-2t}g^{a_1\cdots a_n}(t)(\g_{a_1})^{\a_1\b_1}\cdots (\g_{a_n})^{\a_n\b_n}y^-_{\a_1} y^-_{\b_1}\cdots y^-_{\a_n} y^-_{\b_n}\,.
\eeq{gs}
We should mention that expressing $p_\mu$ in terms of the twistor variables $(y^-_\a,y^{+\a})$ is one-to-one only for light-like momenta. However, this does not mean that the momenta in \re{sc1} are restricted to be light-like since we merely use the twistors to represent the action of the dilatation operator on $ g^{a_1\cdots a_n}$. For $g^{(n)}(t)$ the l.h.s. of \re{p1} is then equivalent to
\beq
D_t g^{(n)}(t)\equiv\partial_t g^{(n)}(t)+[D,g^{(n)}(t)]_*
\eeq{rgweyl}
In particular, the linearized RG-equation for this class of couplings around a fixed point becomes $D_t g^{(n)}(t)=0$.

%
\section{Linearized RG flow as HS equation on AdS}
In this section we will identify $D_t g^{(n)}(t)=0$ with the linearized HS equation of motion on $AdS_4$. To do so we first express $P_{\a\b}$ in terms of $4$-dimensional spinor variables through
\cite{Vasiliev:2012vf} 
\ber
P_{\a\b}=iy^-_\a y^-_\b&=&\frac{1}{4}(i\bar y_\a + y_\a)(\bar  y_\b -i y_\b)\\
&&\to \frac{i}{4}\bar y_{\dot \a}\bar y_{\dot \b}- \frac{i}{4} y_{ \a} y_{ \b}+ \frac{1}{4}\bar y_{\dot \a} y_{ \b}+ \frac{1}{4}y_{ \a} \bar y_{\dot \b}\,.\non
\eer{}
Next we express $D$ in terms of $4$-dimensional spinor variables,
\ber
D=\frac{1}{2}\e_{\a\b}y^{+\a}y^{-\b} &=& \frac{1}{8}\e_{\a\b}( y^{\a} -i \bar y^{\a}) (\bar y^{\b} -i y^{\b})= \frac{1}{4}\e_{\a\b}y^{\a} \bar y^{\b}\non\\
  &&\to  -\frac{i}{4}(\s_2)_{\a\dot\b}y^{\a}\bar y^{\dot\b} 
  \eer{D4}
We now want to compare \re{rgweyl} with the HS equation \cite{Vasiliev:1992av}
\beq
dC+W*C-C*W=0
\eeq{vc}
where $W=W_\m dx^\m$ is an HS gauge potential and $C$ is the field which we will shortly 
identify with the coupling constants describing perturbations of the RG-fixed point. The integrability condition for \re{vc} reads  $dW+W* W=0$. The form \re{D4} of the dilatation operator $D$ picks the  $AdS_4$ solution to this latter equation. Indeed,  the  $AdS_4$ solution is given by 
\beq
W_0(x|Y)=e_0(x|Y)+ \omega_0(x|Y)\,,
\eeq{ads1}
where (in Poincar\'e coordinates)
\beq
e_0(x|Y)=\frac{1}{4i}\frac{dx^\m}{z}(\s_{\m})_{\a\dot\b}y^\a\bar y^{\dot\b}\,,\quad \omega_0(x|Y)=-\frac{1}{4i}\frac{dx^i}{z}((\s^{iz})_{\a\b}y^{\a} y^{\b}+(\bar\s^{iz})_{\dot\a\dot\b}\bar y^{\dot\a}\bar y^{\dot\b})\,.\non
\eeq{}
Our conventions for the 4D sigma matrices are $(\s_\m)_{\a\dot\b}=(1,\vec{\s})_{\a\dot\b}$, $(\bar \s_\m)^{\dot \a\b}=(1,-\vec{\s})^{\dot \a\b}$. The 3D gamma matrices are then obtained by deleting the matrix with space-time index $\m=2$, ie. 
 $(\g_a)_{\a\b}=(1,\s_1,\s_3)_{\a\b}$. Identifying $t=\ln z$ we then see that expression \re{D4} for D is identical with $(W_0)_t(x|Y)$. Thus the tree-level, linearized RG-flow (l.h.s. of \re{p1}) is identified with the HS equation \cite{Vasiliev:1992av}
 \beq
 \partial_t C(x|Y)+[(W_0)_t(x|Y),C(x|Y)]_*=0\,,
 \eeq{vl1}
 where $C(x|Y)$ is given by $g^{(n)}(t)$ is as in \re{gs} but  expressed in terms of 4-dimensional spinor variables using
 \ber
4 (\g^{a})^{\a\b} y^-_{\a} y^-_{\b}   & =&(\g^{a})^{\a\b} (\bar y_{ \a}\bar y_{\b}-  y_{ \a} y_{ \b}-i\bar y_{ \a} y_{ \b}-iy_{ \a} \bar y_{\b})\non\\
 &=&-i(\g^{a})^{\a\b} (\bar y_{\a} y_{ \b}+y_{ \a} \bar y_{ \b})+\e^{abc}(\g_{bc})^{\a\b} (\bar y_{ \a} \bar y_{ \b}-y_{ \a}  y_{ \b})\non\\
 &\to& -i(\s^{a})^{\a\dot\b} y_{ \a} \bar y_{\dot \b} -i(\s^{a})^{\dot \a\b} \bar y_{ \dot \a}  y_{ \b}+\e^{abc}\left((\bar\s_{bc})^{\dot \a\dot\b} \bar y_{\dot \a} \bar y_{\dot \b}-(\s_{bc})^{ \a\b} y_{ \a}  y_{ \b}\right)\non\\
 &=&-2i(\s^{a})^{\a\dot\b} y_{ \a} \bar y_{\dot \b} +2i\left((\bar\s^{za})^{\dot \a\dot\b} \bar y_{\dot \a} \bar y_{\dot \b}+(\s^{za})^{ \a\b} y_{ \a}  y_{ \b}\right)\,,
 \eer{py4}
and we defined $\g_{ab}=\frac{1}{4}[\g_a,\g_b]$. 
In order to verify the other components of the equation \re{vc}, i.e. 
 \beq
 \partial_a C+(W_0)_a*C-C*(W_0)_a=0
 \eeq{rvl}
we then observe that \re{py4} is proportional to $(W_0)_a$ in \re{ads1}. This then implies that \re{rvl} is satisfied so long $C(x|Y)$ does not depend on $x^a$, ie. $C(x|Y)=C(t|Y)$. 

To summarize, {\it eqn. \re{vl1} is the tree-level linearized RG-equation for the couplings $g^{(n)}$ around the Gau\ss ian  fixed point while \re{rvl} encodes translation invariance}. 

Note that in   \re{vl1} and \re{rvl} the commutator $[(W_0)_a, C]_*$  rather than the twisted commutator  
$(W_0)_a*C-C*(\tilde W_0)_a$ with $\tilde f(x|y,\bar y)\equiv  f(x|-y,\bar y)$ enters in the equation for $C(x|Y)$. 
This means that $C(x|Y)$ is a non-propagating (auxiliary) field. This is to be expected since the RG-equation is a first order rather than a second order equation. In other words, the auxiliary sector is responsible for couplings (moduli) of the theory. A similar observation was made in 3-dimensional HS theory considered in \cite{Prokushkin:1998bq}  which describes HS interactions of 3d matter fields of an arbitrary mass. In this case, as explained in that paper, the mass parameter 
is directly related to the value of some of the auxiliary fields. 

To make the relation between HS- and RG-equations complete we should also consider the Vasiliev equations involving the  auxiliary spinor connection, $ S(Z,Y, k,\bar k)=s_\a dz^\a+s_{\dot\a}dz^{\dot\a}$, that is  \cite{Vasiliev:1992av}
\begin{eqnarray}
S*S&=&-i dz_\a dz^\a(1+F(B)*\kappa)-idz_{\dot\a}dz^{\dot\a}(1+\bar F(B)*\bar\kappa)\non\\
S*B-B*S&=&0 \non \label{av}\\
d S+W*S-S*W&=&0
\end{eqnarray}
Here, $Z=(z^\a, z_{\dot\a})$ is pair of auxiliary twistor variable $z^\a$ with $[z^\a,z^\b]_*=-2i\e^{a\b}$ and $ dz^\a dz^{\dot\a}=-dz^{\dot\a}dz^{\a}$ are anti-commuting differentials. The field $B(x|Y,Z)$ is such that $B(x|Y,Z)|_{Z=0}= C(x|Y)$ and otherwise determined by the Vasiliev equations. $F(B)$ is so far an arbitrary function and $\kappa=k K$ where $k$ is a Kleinian with the property
\beq
kf(z^\a,dz^\a,y^\a, z^{\dot \a},dz^{\dot \a},y^{\dot\a})=f(-z^\a,-dz^\a,-y^\a, z^{\dot \a},dz^{\dot \a},y^{\dot\a})k
\eeq{dw}
and  $K=e^{iz_\a y^\a}$ is an inner Kleinian for the $*$-product with the properties
  \beq
  f(y,z)*K=f(-z,-y)K\,,\hbox{and}  \qquad K*f(y,z)=Kf(z,y)
  \eeq{K}
To 0-th order in $B$, the first equation in \re{av} is solved by $S_0=z_\a dz^\a+z_{\dot \a}dz^{\dot \a}$. With this second equation implies that $B$ is independent of $Z$ at leading order (which is the case). To 0-th order in $B$ the last equation is then again identically satisfied. On the other hand, for non-vanishing $F(B)$ this equation implies a correction to $W$ to which we will return below. 

\section{Gauge Transformations}
A complete understanding of how to represent arbitrary HS gauge transformations in RG equations is still lacking. We can nevertheless develop  some intuition by considering examples. The HS gauge transformations  are of the form \cite{Vasiliev:2012vf}
\ber
 \d B(x|Y,Z)&=&-\e(x|Y,Z)*B(x|Y,Z)+B(x|Y,Z)*\e(x|Y,Z)\,.\label{gtbw}\non\\
 \d W(x|Y,Z)&=& d\e(x|Y,Z)+[W,\e]_*(x|Y,Z)
\eer{}
Let us first assume that $\e=\e(t|y^\a,\bar y^{\dot\a})$. In that case the first line in  \re{gtbw} reduces to 
\beq
\d g^{(n)}=[g^{(n)},\e]_*\,
\eeq{gt2}
In order to obtain a 3-dimensional interpretation of $\d g^{(n)}$ we should express $y^\a$ and $\bar y^{\dot\a}$ in terms of $y^{\pm\a}$. For instance, if we take $\e=b^a(t)K_a$ where  $K_a$ is given by \re{ls} expressed in 4D spinor variables, this then amounts to
\beq
g^{(n)} \to g^{(n)}+2(g\cdot b)^{(n-1)} D+2i(g^{[a}b^{b]})^{(n-1)}L_{ab}\,,
\eeq{dgk}
which can then be realized in field theory as lower dimensional, momentum dependent couplings in $\cS_t$. In \re{dgk} symmetric (Weyl) ordering is understood. This transformation does not leave $\cS_t$ invariant, generically\footnote{It is conceivable that an invariant action can nevertheless be found along the lines  \cite{Bekaert:2009ud} where a consistent coupling of a scalar to higher spin external fields is constructed.}
, although some terms may vanish. For instance, for $n$ odd $\cS_t$ does not not depend on $g^{(n)}$ as we already noted. Similarly, the Lorentz term on the r.h.s. of \re{dgk} vanishes again in $\cS_t$ for $n$ odd while the dilatation term does contribute a coupling with $n$ even. The connection, on the other hand, transforms as
%
\beq
D\to D+\partial_tb^aK_a-b^aK_a
\eeq{dD}
Clearly, $W_0+\d W$ is no longer of the $AdS$-form, not even modulo coordinate transformations which are parametrized by $\e(x)$.  Nevertheless,  \re{rgweyl} transforms covariantly by construction. Explicitly, 
\beq
 [D,\d g^{(n)}]+[\d D, g^{(n)}]=2n(\d g)^{(n-1)}-2(\dot b\cdot g)^{(n-1)} D-2i(\dot b^{[a} g^{b]})^{(n-1)}L_{ab}
 \eeq{}
 and 
 \beq
 \partial_t \d g^{(n)}=2(\dot b\cdot g)^{(n-1)} D+2i(\dot b^{[a} g^{b]})^{(n-1)}L_{ab}+2 (b\cdot\dot g)^{(n-1)} D+2i( b^{[a} \dot g^{b]})^{(n-1)}L_{ab}\,.
 \eeq{}
 Since $\dot g^{(n)}=-ng^{(n)}$, \re{rgweyl} is satisfied without using further properties of $\cS_t$. 
 If $b(t)$ satisfies the RG-equation then $\d D$ in the connection $D$ in \re{dD}  is invariant  so that \re{rgweyl} in its original form is valid for $  g^{(n)}$ as well as $\d g^{(n)}$.  
 It is interesting to note that the RG equation displays an $o(3,2)$ even if the action parametrized by $\{g^{(n)}\}$ is not conformal. Of course, we have only considered the linearized flow so far. 
 
 As an example of a $z$-dependent gauge transformation we take $\e=K$ as in \re{K}. Then $\d B=-K*B+B*K$ or, equivalently
\beq
\d g^{(n)}=-K*g^{(n)}+g^{(n)}*K=-K(z,y)g^{(n)}(z^\a,\bar y^{\dot\a})+g^{(n)}(-z^\a,\bar y^{\dot\a})K(z,y)
\eeq{}
Remember that $g^{(n)}$ was independent of $z^\a$ in our construction but this is not a gauge-invariant statement. Since there is no interpretation for $z^\a$ in 3-dimensions I suspect that the correct interpretation in 3 dimensions is to set $z^\a=0$. Thus $\d g^{(n)}=0$.


\section{Adding the Inhomogeneity}
Let us finally return to the r.h.s. of \re{p1}. In \cite{Douglas:2010rc} it was suggested to include this terms in a redefinition of $W_0$. However, we shall see below that this is realized differently in the HS description. For a free field theory the quantum term on the r.h.s. can only contribute to the cosmological constant which we already discussed and which is not part of the higher spin spectrum. 
Let us now turn to the first (tree-level) term on the r.h.s. of  \re{p1} which corresponds to the graph in fig. 1a.  
This graph contributes a source term of the form 
\beq
\int \vf(-p)g^{a_1\cdots a_m}g^{b_1\cdots b_n}\cK'(p^2)p_{a_1}\cdots p_{a_m}p_{b_1}\cdots p_{b_n}\vf(p)
\eeq{pc}
to the r.h.s. of  \re{p1}\footnote{ In $x$-space this sources a non-local coupling through
\ber
 \partial_t g^{a_q\cdots a_{n+m}}(x)(\partial_{a_1}\cdots  \partial_{a_{m+n}}\varphi(y))\varphi(x+y)= g^{a_1\cdots a_{n}}g^{a_{n+1}\cdots a_{n+m}}\cK'(x)(\partial_{a_1}\cdots  \partial_{a_{m+n}}\varphi(y))\varphi(x+y)\,\nonumber
\eer{nl1}
Taking the symmetrized, traceless part of both sides this amounts to 
\beq
D_t g^{(n+m)}(t,x)=\cK'(x)e^{-2t}g^{a_1,\cdots, a_n}(t)g^{a_{n+1},\cdots, a_{n+m}}(t)\g_{a_1}^{\a_1\b_1}\cdots \g_{a_{n+m}}^{\a_{n+m}\b_{n+m}}y^-_{\a_1} \cdots y^-_{\b_{n+m}}\non\,.
\eeq{nltl}}
Of course, the product of two traceless couplings as in (\ref{pc}) need not be traceless so that traceful couplings will be generated along the RG-flow. In order to clarify the algebraic structure it is then useful to define a product "$\cdot$" on the space of couplings  through
\beq
g^{a_1\cdots a_m}\cdot g^{b_1\cdots b_n}=g^{(a_1\cdots a_m}g^{b_1\cdots b_n)}
\eeq{pdot}
where $(\cdots)$ stands for symmetrization of the indices. This product is clearly associative. If we furthermore divide out the ideal generated by elements containing traceful couplings (corresponding to deformations of  $\cS_t$ that involve the d'Alembertian, $p^2$) then the set of traceless symmetric higher derivative couplings with the above product  form an abelian algebra\footnote{If we allow the couplings to depend on the coordinates of the 2+1 dimensional field theory this algebra is no longer abelian. In that case, upon suitable ordering of the derivatives as in \cite{Bekaert:2009ud} the couplings generate the higher spin algebra.}.

In order to describe the contribution from  (\ref{pc}) to the flow of the traceless couplings we expand $\cK'$ as  $\cK'(p^2)=\cK'(0)+O(p^2)$. Then, recalling that representation of the momenta $p_a$  in term of spinor variable as in (\ref{pd}) takes care of symmetrization and projection onto vanishing trace automatically the non-linear correction to (\ref{vl1}) can be written as 
 \beq
 \partial_t C(Y|t)+[(W_0)_t(t|Y),C(Y,t)]_*=2\cK'(0)C(Y,t) C(Y,t)\,.
 \eeq{vl1i}
 As it stands this equation does not look covariant under the higher spin gauge transformations (\ref{gtbw}) not even when restricted to $\e(x|Y,Z)$ that are independent of $Z$. Note, however that we may as well replace $C(Y,t) C(Y,t)$ by the star products $C(Y,t)*C(Y,t)$ since on functions that depend only on $y_\a^-$  both products agree. Thus (\ref{vl1i}) is, in fact covariant under the higher spin gauge transformations. Finally, the traceful couplings sourced by  (\ref{pc}) and sitting  in the ideal defined above, do not "back react" on the traceless couplings at tree-level. 
 
  Let us now analyze what happens in the presence of interactions. Here we focus on the quartic term  $\lambda \int \vf(p_1)\cdots\vf(p_4)\delta^3(p_1+\cdots+p_4)$ in $\cS_t$ which is interesting in connection with the $O(N)$ model. At tree-level, the RG-flow  \re{p1} produces, among higher order couplings, also traceless couplings of the form ( fig. 1b)  
  \beq
  2\lambda\cK'(0)g^{a_1\cdots a_n}\int  q_{a_1}\cdots q_{a_n}\vf(q)\vf(p_2)\cdots\vf(p_4)\delta^3(q+\cdots+p_4)
  \eeq{lg}
which are not captured by the minimal higher spin system  (\ref{vl1i}) but, again do not effect the running of the traceless, quadratic couplings at tree-level  and thus the system (\ref{vl1i}) is indeed closed at tree-level.
 
 At quantum level ($2nd$ term on the r.h.s of  \re{p1}) the above decoupling no-longer takes place. Indeed, at one-loop a typical contribution to the flow of $g^{a_1\cdots a_n}$ is displayed in ( fig. 1b) and, up to some numerical factors, gives a contribution of the form
 \beq
g^{a_1\cdots a_n} \lambda\cK'(0)(g^{(s)},g^{(s)})\int  (p^2)^s\cK'(p^2)
 \eeq{lgg}
 where $(g^{(s)},g^{(s)})=g^{a_1\cdots a_s}g_{a_1\cdots a_s}$ is an element in the ideal. Such contributions to the flow can be incorporated by noticing that they contribute to the anomalous dimension $\gamma (g^{(s)},\lambda)$ as a HS-singlet. They are consistent with our HS equation  (\ref{vl1i}) provided modify the definition  (\ref{gs}) as 
\beq
g^{(n)}(t)\equiv e^{-2(1+\gamma (g^{(s)},\lambda))t}g^{a_1\cdots a_n}(t)(\g_{a_1})^{\a_1\b_1}\cdots (\g_{a_n})^{\a_n\b_n}y^-_{\a_1} y^-_{\b_1}\cdots y^-_{\a_n} y^-_{\b_n}\,.
\eeq{gsn}
Consistency of   (\ref{vl1i})  follows from the fact that $\gamma (g^{(s)},\lambda)$ is invariant under HS symmetry transformations. Higher order corrections to $\gamma$ in $\lambda$ arise among others from the "bubble diagrams" (see e.g. \cite{Moshe:2003xn}) 

Note that all diagrams discussed so far and that contribute to the running of $g^{(n)}$ are proportional to $\cK'(0)$. This implies, in particular that, if we choose a cut-off function such that $\cK'(0)=0$, then none of these diagrams contribute to the running of $g^{(n)}$ for $n>0$. The scalar coupling is an exception since there is a contribution to the running of the mass $g^{(0)}$ due to standard mass renormalization in $\vf^4$-theory. Of course, there are many more diagrams that contribute to the running of $g^{(n)}$ even for  $\cK'(0)=0$. One such diagram is depicted in 
fig. 1d. While this latter contribution can be absorbed in the anomalous dimension as in (\ref{lgg}) it is not clear that this is the case for all higher diagrams of this type. 

A substantial simplification occurs if we consider the large $N$ limit of this vector model with Lagrangian (see e.g. \cite{Moshe:2003xn}) 
\beq
\cL=  \frac{1}{2} \pa_\a\vf^i\pa^\a\vf^i+ \frac{1}{2}m^2\vf^i\vf^i+ \frac{\lambda}{4!N}(\vf^i\vf^i)^2
\eeq{nv}
Contributions of the form fig. 1d are then suppressed in the large N limit. Thus we conclude that 

{\it In the large N limit the RG-equations for the traceless higher derivative couplings $g^{(n)}$ are precisely given by  (\ref{vl1i}) safe the mass term  $g^{(0)}$. Furthermore, for $\cK'(0)=0$ equation (\ref{vl1i}) reduces to the classical equation (\ref{vl1}).}

It is reassuring to note that the higher spin couplings $g^{(n)}$, $\geq 2$ are (marginally) irrelvant in the IR so that the IR fixed point is independent of the value of $\cK'(0)$. 

\section{Discussion}
In this paper we found that the classical RG-flow for traceless higher derivative couplings that are quadratic in the fields is an interacting HS-equation of motion for the auxiliary, non-propagating  sector on an AdS background. In the large N limit of an N-component interacting scalar field theory this equivalence extends to the full quantum theory even away from the conformal fixed points. The fact the auxiliary sector is related to couplings in the RG-equation seems physically sensible since the RG-flow is a first order flow. Although our derivation was done for a scalar field in 2+1 dimensions other types of fields can be treated on same footing. Generalizations to other space-time dimensions should also be possible but the 3-dimensional case is particularly intuitive due to the simple relationship between 3- and 4-dimensional twistor variables. An interesting question is whether the relation found here can be extended to finite values of N. 

Another interesting question is whether the HS-nature of the RG-equation is useful in explaining the origin of the HS/O(N)-model duality which involves the physical, propagating sector of the Vasiliev theoy. Perhaps progress can be made in combining the map constructed her here with previous ideas developed in \cite{Bekaert:2010ky} for instance. One possible hint comes from the observation that the auxiliary HS-modes which were related to the couplings in this paper source physical HS-modes on AdS once the interacting Vasiliev theory is considered \cite{Vasiliev:1992av}. From our point of view it is therefore natural to interpret the latter modes as vevs.  We hope to come back to this issue in the future. 

\vspace{1cm}
\begin{figure}
%
\hspace{0.9cm}
\begin{tikzpicture}
  \draw (-0.5,.1) -- (0.4,0.1);
   \draw[line width=1mm] (0.3,.1) -- (1.3,0.1);
     \draw (1.6,0.1) -- (2.0,0.1);
 \node at (0,.1)  [rectangle,draw=black!50,fill=black!20] {$n_1$};\hspace {0.7cm}
\node at (0.8,.1)  [rectangle,draw=black!50,fill=black!20] {$n_2$};
\end{tikzpicture}
\hspace{1.0cm}
\begin{tikzpicture}
  \draw (-0.5,.1) -- (0.4,0.1);
   \draw[line width=1mm] (0.3,.1) -- (0.7,0.1);
     \draw (0.7,0.1) -- (1.4,0.1);
     \draw (0.7,0.1) -- (1.3,0.5);
     \draw (0.7,0.1) -- (1.3,-0.3);
 \node at (0,.1)  [rectangle,draw=black!50,fill=black!20] {$n$};\hspace {0.7cm}
\end{tikzpicture}
\hspace{-0.2cm}
\begin{tikzpicture}
%
 \draw (-0.5,0) -- (2.5,0);
  \draw[line width=1mm] (0.7,0) -- (1.5,0);
 \tikz \draw (0,0) circle (20pt);
\node at (-0.1,0.7)  [rectangle,draw=black!50,fill=black!20] {s};
\node at (-1.4,0.7)  [rectangle,draw=black!50,fill=black!20] {s};
 \node at (0.3,0)  [rectangle,draw=black!50,fill=black!20] {n};
\end{tikzpicture}
\hspace{1.6cm}
\begin{tikzpicture}
%
  \draw(-0.7,0.7) -- (2.3,0.7);
 \tikz \draw (0,0) circle (20pt);
\node at (-0.7,0.7)  [rectangle,draw=black!50,fill=black!20] {n};
\end{tikzpicture}

 \caption{\quad $(a)$\qquad\qquad\qquad $(b)$ \qquad\qquad\qquad\quad$(c)$\qquad \qquad\qquad\qquad $(d)$ \newline  \newline Classical- and quantum contributions the r.h.s. of \re{p1}. A fat line represents an insertion of $2\cK'(0)$ and the squares represent traceless, quadratic higher derivative couplings,  $g^{(n)}$.}
  \label{fig1}
\end{figure}
%
%

\noindent{\bf Acknowledgements:}
I would like to thank O. Hohm for early collaboration and D. Ponomarev for many helpful discussions throughout this work, M. Vasiliev for his comments on an earlier version of this note as well as N. Boulanger, A. Rovin and S. Konopka for helpful discussions. This project was supported in parts by the DFG Transregional Collaborative Research Centre TRR 33, the DFG cluster of excellence ÒOrigin and Structure of the UniverseÓ, the DFG project HO 4261/2-1 as well as the DAAD project 54446342.

\end{document}